\documentclass[preprint]{elsarticle}

\usepackage{spverbatim}
\usepackage{listings}
\usepackage{url}
\usepackage{hyperref}
\usepackage[]{algorithm2e}
\usepackage{xcolor}

\usepackage{comment}

\usepackage{subcaption}
\usepackage{lipsum} 

\usepackage{float}

\begin{document}
\title{Leveraging AI to optimize website structure discovery during Penetration Testing}

\author[2]{Diego Antonelli}
    \ead{diego.antonelli@nttdata.com}
\author[2]{Roberta Cascella}
    \ead{roberta.cascella@nttdata.com}
\author[1]{Gaetano Perrone}
    \ead{gaetano.perrone@unina.it}
\author[1]{Simon Pietro Romano}
    \ead{spromano@unina.it}
\author[2]{Antonio Schiano}
    \ead{antonio.schiano@nttdata.com}
    
\address[1]{University of Napoli Federico II}
\address[2]{NTT DATA Italia S.p.A.}

\begin{abstract}
Dirbusting is a technique used to brute force directories and file names on web servers while monitoring HTTP responses, in order to enumerate server contents. 
Such a technique uses lists of common words to discover the hidden structure of the target website. Dirbusting typically relies on response codes as discovery conditions to find new pages. It is widely used in web application penetration testing, an activity that allows companies to detect websites vulnerabilities. Dirbusting techniques are both time and resource consuming and innovative approaches have never been explored in this field. We hence propose an advanced technique to optimize the dirbusting process by leveraging Artificial Intelligence. More specifically, we use semantic clustering techniques in order to organize wordlist items in different groups according to their semantic meaning. The created clusters are used in an ad-hoc implemented next-word intelligent strategy. This paper demonstrates that the usage of clustering techniques outperforms the commonly used brute force methods. Performance is evaluated by testing eight different web applications. Results show a performance increase that is up to 50\% for each of the conducted experiments.
\end{abstract}

\begin{keyword}
Artificial Intelligence \sep Dirbusting \sep Network Security \sep Penetration Testing \sep Performance Assessment \sep Semantic Clustering.
\end{keyword}

\maketitle

\section{Introduction} 
\label{section:intro}
{W}{eb} application penetration testing is an offensive approach to find web vulnerabilities. It is a process that allows companies to find vulnerabilities in websites, by using a black-box approach. A Web Application Penetration Test (WAPT) process is composed of different phases: \emph{information gathering}, \emph{enumeration}, \emph{exploitation} and \emph{analysis}.
Information Gathering is aimed at exploring the web application with the purpose of finding information about used libraries and frameworks, as well as to understand the structure of files and folders.
OWASP (Open Web Application Security Project\footnote[1]{\url{https://owasp.org/}}) defines a standard methodology to test the security of a Web Application.  The OWASP Security Testing methodology \cite{williams2006owasp} defines different security categories (four characters upper case strings) that identify the type of test or weakness. Each category contains different Security Tests, indexed by a digit. 
According to the OWASP Methodology, discovering the structure of a web application is an important activity included in the \textit{``Fingerprinting Web Application'' Security Test (WSTG-INFO-09)}.
Each web application has its own specific files and folders structure. Since Web Application Penetration Test is a black-box activity, the penetration tester has no a priori knowledge about the structure of the web application. 
Indeed, different techniques exist for the discovery of available paths at the server. As an example, \emph{spidering} analyzes internal links inside HTML pages and navigates through them to discover both files and folders. Though, links that are not directly present in the HTML source code often exist. For this reason, penetration testers use other techniques to discover the directory structure of a web application. One such technique is called \emph{dirbusting} and will be briefly described below.

\subsection{Dirbusting}
Dirbusting is a technique used to brute force a target with predictable folder and file names while monitoring HTTP responses to enumerate server contents. 
This technique uses wordlists to send HTTP requests to a target website and discover hidden pages. It is useful during the first phase of a Penetration Testing activity in order to discover the structure of the target application. It is important to remark that Penetration Testing is usually carried out as a black-box activity. The security expert has no access to information about the web application under test, and typically has just low privileged access to the system. For this reason, she/he is not able to see all the pages of a web application by just using spidering. Thus, one of the goals of dirbusting is to discover pages that are not visible by using common spidering techniques. Hidden pages might allow a security expert to find sensitive content on the website, or valid entry-points to perform other vulnerability injection tests. Dirbusting accepts as input a properly constructed list of words and starts sending HTTP requests to the website to discover new pages. In order to successfully complete its task, the dirbusting process needs a proper discovery condition. A common approach is to use the response code for that purpose: a new page is found when an HTTP response contains a status code other than 404 (Page Not Found). In this work, we define  ``valid requests'' those HTTP requests that have a response code other than 404. To obtain good results in terms of discovered pages, the choice of wordlists plays a crucial role.
Such a choice depends on the acquired knowledge about the web application. Security experts choose the wordlists on the basis of several criteria, like, e.g.:

\begin{itemize}
    \item which convention the developer has used to define paths;
    \item which framework/CMS (Content Management System) has been used;
    \item which language has been adopted to develop the web application.
\end{itemize}

If the web application contains a page whose name comes with camel case notation (e.g., \emph{loginPage}), it is advisable to use camel case wordlists (\emph{logoutPage}, \emph{adminPage}, etc.). Similarly, if the fingerprinting phase detected the existence of a Wordpress Content Management System, an optimized wordlist should contain Wordpress-specific words (\emph{wp-login.php}, \emph{wp-logout.php}, etc.). On the other hand, if the web application contains files with well-known extensions (e.g., JSP, PHP), it is better to use a wordlist whose stems properly fit them. 

In this work, we demonstrate that a \emph{semantic clustering} strategy is able to optimize dirbusting activities by properly mimicking the behavior of a security expert when it comes to choosing the most appropriate wordlist. Before delving into the details of the proposed approach, we will hence briefly introduce semantic clustering in the next subsection.

\subsection{Semantic Clustering}

Clustering is the process of partitioning a set of data objects into subsets in such a way that items in the same group are more similar to each other than to those in other groups. 
The objective is to maximize intra-cluster similarity while at the same time minimizing inter-cluster similarity. It is a widely used technique in data mining for text domains, where the items to be clustered are textual and they can be of different granularity (documents, paragraphs, sentences or terms).

Simple text clustering algorithms represent textual information as a document-term matrix. Features are computed based on term frequencies and semantically related terms are not considered. Thus, documents clustered in this way are not conceptually similar to one another if no terms are shared, as semantic relationships are ignored. 
 
 Semantic clustering, instead, consists in grouping items into semantically related groups~\cite{ibrahim2019survey}\cite{naik2015survey}. This requires to measure the semantic similarity between textual information, which can be accomplished by vectorizing the text corpus using, among the others, one of the following resources:
\begin{itemize}
    \item  Semantic networks like WordNet~\cite{fellbaum2012wordnet}: a large lexical database of more than 200 languages. Nouns, verbs, adjectives and adverbs are grouped into sets of cognitive synonyms (synsets), each expressing a distinct concept. Synsets are interlinked by means of conceptual-semantic and lexical relations.
    
    \item Word embeddings techniques such as Word2Vec~\cite{mikolov2013efficientw2v1}\cite{mikolov2013distributedw2v2} and GloVe~\cite{pennington2014glove}: a word embedding is a simple neural network trained to reconstruct the linguistic contexts of words.
Its input is a large corpus of words and produces a vector space, with each unique word assigned to a corresponding vector.
Word embeddings give us a way to use an efficient, dense representation in which similar words have a similar encoding.

    \item  Sentence Embeddings: while word embeddings encode words into a vector representation, sentence embeddings represent a whole sentence in a way that a machine can easily work with. These are capable of encoding a whole sentence as one vector. Examples are Doc2Vec~\cite{le2014distributed}, an adaptation of word2vec for documents, or more recent approaches such as the Universal Sentence Encoder (USE)~\cite{cer2018universal} and InferSent~\cite{conneau2017supervised}.
    
    \item  Language representation models like the Bidirectional Encoder Representations from Transformers (BERT)~\cite{devlin2018bert}: BERT is a method of pre-training language representations, meaning that a general-purpose ``language understanding'' model is trained on a large text corpus (e.g. Wikipedia), and then used for downstream Natural Language Processing tasks. Pre-trained representations can either be context-free or contextual. Context-free models such as word2vec~\cite{mikolov2013efficientw2v1}\cite{mikolov2013distributedw2v2} or GloVe~\cite{pennington2014glove} generate a single word embedding representation for each word in the vocabulary, so, for example, the word \textit{basket} would have the same representation in \textit{sports} and \textit{e-commerce}. Contextual models, instead, generate a representation of each word that depends on the other words in the sentence.
\end{itemize}

  Our approach leverages the Universal Sentence Encoder (USE)~\cite{cer2018universal} as the chosen sentence embeddings technique. Indeed, one of the main tasks for training a USE encoder is the identification of the semantic textual similarity (STS)~\cite{cer-etal-2017-semeval} between sentence pairs scored by Pearson correlation with human judgments. A task that perfectly fits our needs.

After the text corpus encoding phase, it is required to use a clustering algorithm in order to create the semantic clusters. There are several clustering techniques which can be effective for this purpose~\cite{jain2010data}, and among the available choices, the K-means algorithm is used for its simplicity and accuracy. 

\begin{figure}
    \centerline{
    \includegraphics[width=0.8\linewidth]{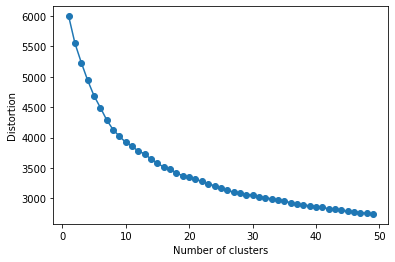}}
    \caption{Elbow method}
    \label{fig:elbowMethod}
\end{figure}

One of the issues with K-means is the effective choice of the parameter \textit{K}, i.e., the number of target clusters. In our case, such an issue is solved by leveraging the well-known \emph{elbow method}, a heuristic used in determining the number of clusters in a data set. Fig.~\ref{fig:elbowMethod} provides a graphical representation of such a heuristic.

We demonstrate in this paper that the method we propose allows to improve dirbusting techniques by leveraging artificial intelligence. The approach was tested on 8 different web applications, with 30 repetitions each, showing a substantial performance improvement in each of them.

\section{Related Works}
To the best of our knowledge, the idea of leveraging Artificial Intelligence has never been explored in the field of dirbusting. Relevant works thus fall in the wider area of semantic clustering methodologies, that has been extensively explored in other application domains~\cite{ibrahim2019survey}\cite{naik2015survey}.
Regardless of their application to dirbusting, Natural Language Processing techniques have been extensively used in the security field. Karbab \cite{karbab2019maldy} uses Natural Language Processing and machine learning techniques to create a behavioral data-driven malware detection tool. Malhotra \cite{malhotra2016analyzing} shows that NLP can help evaluate completeness, contradiction and inconsistency of security requirements of a software system.

More in general, the use of Artificial Intelligence techniques for Penetration Testing has not been fully explored yet. Though, several techniques, such as fuzzing, have been used in other domains. As an example, in the software testing field there are several works that show how it is possible to optimize fuzzing techniques by using machine learning \cite{godefroid2017learn}.  Our work rather shows how it is possible to optimize a bruteforce technique (i.e., dirbusting) by using Artificial Intelligence. Hitaj \cite{hitaj2019passgan} shows that a Deep Learning approach is able to outperform both rule-based and state-of-art password guessing approaches.
As password guessing is basically a bruteforce attack, our work shares the idea that it is possible to improve Penetration Testing tasks through AI.  
In particular, Natural Language Processsing techniques can potentially improve tasks that are related to the usage of words. With special reference to dirbusting, semantic clustering can indeed optimize a bruteforce approach by finding both syntactic and semantic relations among words. 


Semantic clusters can be modeled in different ways, including the usage of external resources such as Wikipedia like in~\cite{nourashrafeddin2014ensemble_29}\cite{wu2017efficient_43}, where authors clustered the text corpus with an ensemble approach using knowledge and concept from Wikipedia. Other works~\cite{desai2016wordnet_9}\cite{sahni2014novel17}\cite{wei2015semantic31}\cite{fiorini2016fast_37} leverage semantic networks, such as WordNet~\cite{fellbaum2012wordnet}, which is used  as word sense disambiguation to capture the main theme of text and identify relationships among words. More recent applications use word embedding techniques~\cite{zhang2017automatic}\cite{li2018lda}\cite{alshari2017improvement} and sentence embedding techniques \cite{karagkiozis2019clustering}\cite{asgari2020covid}\cite{bodrunova2020topic} where unsupervised embeddings models are used to encode the text corpus prior to the clustering process.

Our semantic clustering approach follows the works described in~\cite{karagkiozis2019clustering} and~\cite{asgari2020covid}. In~\cite{karagkiozis2019clustering}, Universal Sentence Encoding (USE)~\cite{cer2018universal} and InferSent~\cite{conneau2017supervised} are used to find semantic similarities among user questions and cluster them by using the K-means clustering algorithm. In~\cite{asgari2020covid}, such techniques are instead used to group similar tweets in a semantic sense.

%
%
%
%


\begin{figure*}[t!]
    \centerline{
    \includegraphics[width=\textwidth]{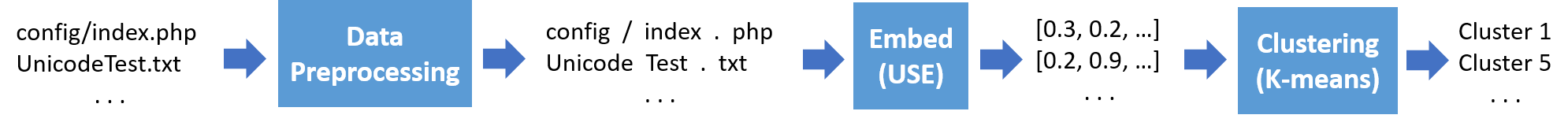}}
    \caption{Semantic Clustering Process}
    \label{fig:EmbeddingsProcess}
\end{figure*}

\section{Proposed Solution}

The proposed approach groups common files and directories contained in the wordlist based on the chosen semantic clustering technique. Semantic clusters define the execution order of the entries in the wordlist, with the aim of optimizing the dirbusting process. Digging more into the details, the approach involves a data pre-processing step on the entries of the wordlist and the subsequent creation of semantic clusters, using the Universal Sentence Encoder (USE), in conjunction with the K-means clustering algorithm, as shown in Fig. \ref{fig:EmbeddingsProcess}.

\subsection{Semantic Clustering}

The first step of the clustering process is data-processing, which consists into splitting each entry of the wordlist according to the naming convention (\textit{camelCase, snake\_case, kabab-case}), and punctuation characters (\textit{! " \$ \# \% \& \' ( ) * + , - . / : ; < = > ? @ [ ] \^ \_ ` \{ | \} \~}). For instance, the sentence \textit{``comments/add\_comment.php''} becomes \textit{``comments  /  add  \_  comment  .  php''} and \textit{``UnicodeTest.txt''} becomes \textit{``Unicode  Test  .  txt''}. This a fundamental step because the naming convention and the punctuation characterizing each entry of the wordlist affect the encoding of the entry itself into embedding vectors. This may lead to a wrong similarity measure.
For this reason, we detect the words contained in each entry, in order to treat them as a sentence instead of a single word. 
In this way, we are able to get the semantic similarity among names contained in a wordlist, regardless of the specific naming convention adopted by the developers.

Then, a sentence embedding technique is used to encode each entry of the wordlist as a $512$-dimensional vector so that similar words, often used in similar contexts, have a similar embedding vectors representation. 
More specifically, the \textit{Universal Sentence Encoder (USE)}~\cite{cer2018universal}, version 4, implemented in TensorFlow 2.2.0~\cite{tensorflow2015-whitepaper} is used. This is a model that encodes text into high-dimensional vectors used for text classification, semantic similarity, clustering and other natural language tasks. The model is trained on a variety of data sources and optimized for greater-than-word length text, such as sentences, phrases or short paragraphs. One of the main tasks for the USE training is the identification of the semantic textual similarity (STS) between sentence pairs, a task that perfectly fits our needs, and hence justifies our choice.


Finally, extracted embeddings are used with clustering techniques to create the semantic clusters. Our approach uses the K-means clustering technique for its simplicity and accuracy. The number of clusters is chosen by using the elbow method, a heuristic used in determining the number of clusters in a data set. The method consists in plotting the explained variation as a function of the number of clusters, and picking a point slightly right to the elbow of the curve as the number of clusters to use, $20$ in this case.

\begin{figure*}
    \centerline{
    \includegraphics[width=0.95\linewidth]{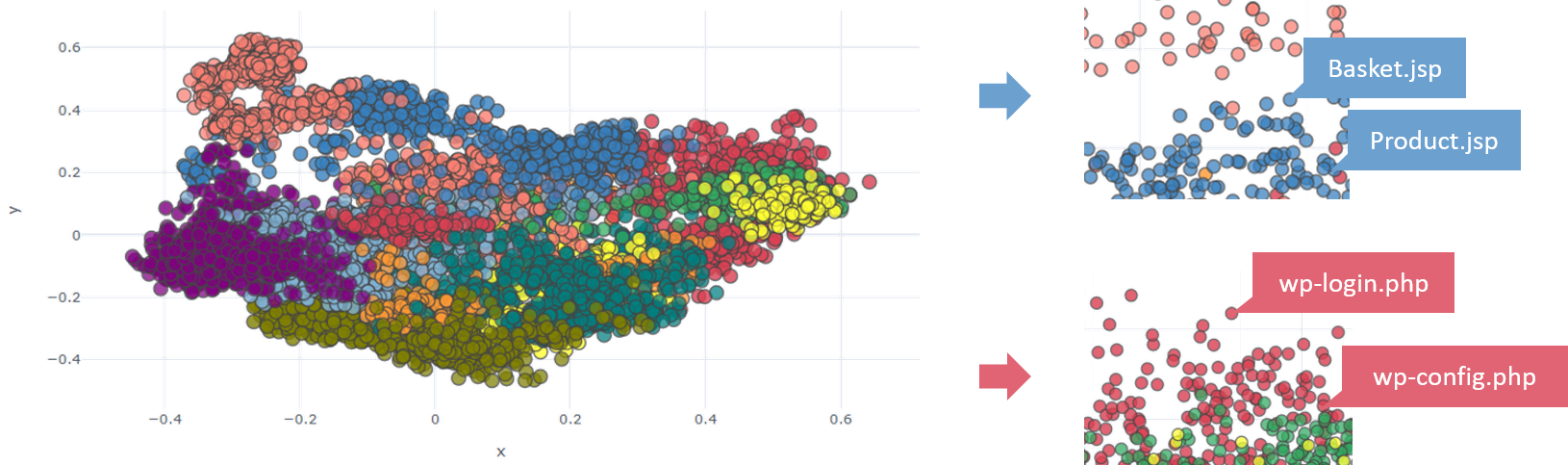}}
    \caption{Semantic Clusters}
    \label{fig:clusters}
\end{figure*}

In Fig. \ref{fig:clusters}, Principal Component Analysis (PCA) is used to show in a two dimensional space the similarities of the words of the wordlist encoded using USE. Each of the points in the picture represents a word (\textit{word\_ik}), and each color represents the cluster (\textit{cluster\_k}) where the word belongs. As shown in the picture, semantically similar words are closer in the embeddings space and are grouped in the same cluster.

Other examples are presented in Table \ref{tab:semanticClustersWordsExample} where words belonging to 5 different clusters are analyzed.

\begin{table}[H]
    \centering
    \begin{tabular}{ | c | c | c |}
    \hline
    Word & Cluster \\ \hline\hline

    libraries/joomla/github/package/gitignore.php & 1   \\ \hline
    plugins/user/joomla/joomla.php & 1   \\ \hline
    libraries/cms/menu/menu.php & 1   \\ \hline
    libraries/cms/component/helper.php & 1   \\ \hline\hline
   
    wp-login.php & 2   \\ \hline
    wp-config.php & 2   \\ \hline
    wp-includes/fonts/dashicons.eot & 2 \\ \hline
    wp-content/plugins/index.php & 2   \\ \hline\hline
    
    about.php & 3   \\ \hline
    appinfo.php & 3   \\ \hline
    index.php & 3   \\ \hline
    update.php & 3  \\
    \hline\hline
   
    basket.jsp & 4   \\ \hline
    product.jsp & 4   \\ \hline
    search.jsp & 4   \\ \hline
    cart/.gitignore & 4   \\ \hline\hline
   
    images/bricks.jpg & 5   \\ \hline
    images/menu/menu\_tabs.gif & 5   \\ \hline
    misc/tree.png & 5   \\ \hline
    favicon.ico & 5   \\ \hline
    
    \end{tabular}
    \caption{Examples of similarities in semantic clustered words}
    \label{tab:semanticClustersWordsExample}
\end{table}

\subsection{Intelligent Dirbusting Strategy}
We implemented an intelligent dirbusting strategy that uses the semantic clusters created according to the proposed approach to improve the legacy brute force techniques.

Commonly used dirbusting techniques require a huge number of requests to the target website, attempting to guess the names or identifiers of hidden functionality based on the common files and directories contained in the wordlists.

The choice of the wordlist depends on the information gathered by the security experts during the spidering process, whose main task is to enumerate the target’s visible content and functionality.
Based on the knowledge acquired during this phase, the experts select the proper wordlist following these criteria:

\begin{itemize}
    \item \textit{Naming conventions}: developers are used to follow a naming convention (camel case, snake case, kebab case) when implementing a web application. For this reason, if a security expert identifies a specific naming convention, the wordlist is chosen or adapted to it.
    
    \item\textit{Content Management Systems (CMS)}: if the fingerprinting phase detected the existence of a certain CMS, a corresponding wordlist is chosen. 
    
    \item \textit{Used programming language}: if the web application under test contains files with well-known extensions (e.g. .JSP ,PHP), it is advisable to use a wordlist whose stems properly fit them.
\end{itemize}

As a general rule, in order to discover as much hidden content as possible, it is fundamental to choose the wordlist that best suits the target's characteristics.
Once the wordlist is chosen, the dirbusting process starts by addressing HTTP requests to the target according to the directory and file names contained in the list itself.
The order of execution of the requests follows the wordlist order, as described in algorithm~\ref{alg: dirbust} below.

\begin{algorithm}

 \While{words in wordlist}{
      pop a word from wordlist (word\_i)\; 
      
      url = basePath + word\_i\;
      
      statusCode = httpRequest(url)\;
      
      \eIf{statusCode is not 404}{
          save word\_i\;
      }{
       continue iterations\;
      }
 }
 \caption{Dirbusting Process}
 \label{alg: dirbust}
\end{algorithm}

\begin{figure}
    \centerline{
    \includegraphics[width=0.5\linewidth]{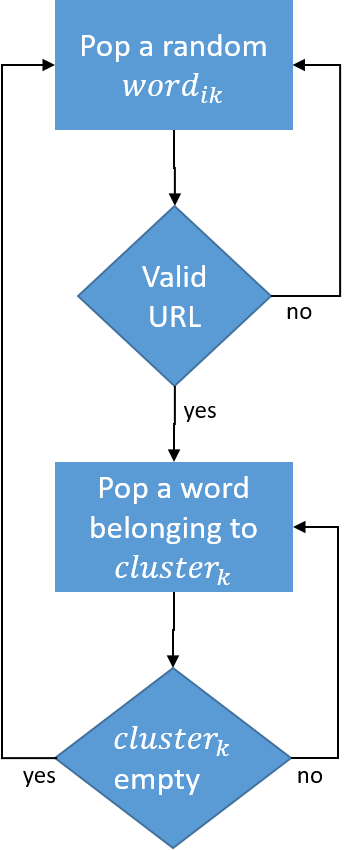}}
    \caption{Semantic Clustering Strategy Flow}
    \label{fig:strategyFlow}
\end{figure}

Our approach aims at making dirbusting more intelligent by leveraging artificial intelligence. As described in the flow in Fig.~\ref{fig:strategyFlow}, the process starts by choosing a random word (\textit{word\_ik}) from a common wordlist. When a valid URL is detected, the proposed strategy consists in choosing the cluster (\textit{cluster\_k}) where the current word belongs to. 
In this way, the next words picked from the chosen cluster will likely target another valid URL. Examples of words grouped in the same cluster are given in Tab.~\ref{tab:semanticClustersWordsExample}.

While common dirbusting techniques require that the expert is forced to manually select a wordlist according to the target characteristics, the intelligent dirbusting strategy accomplishes this task by building the above described semantic clusters, while considering the following aspects:

\begin{itemize}
    \item the CMS used in web applications:  as shown in Table~\ref{tab:semanticClustersWordsExample}, in clusters $1$ and $2$, words related to different CMSs are grouped together. In cluster $1$, Joomla-related words are included, while in cluster $2$ we can find words commonly used in WordPress.
    
    \item web application programming languages: clusters $3$ and $4$, in Table~\ref{tab:semanticClustersWordsExample}, include words related to specific languages. In these clusters, the programming languages \textit{.PHP} and \textit{.JSP}, respectively, are represented.
    
    \item semantic similarities: semantic clusters are able to consider semantic similarities as well. In this way, dirbusting is able to automatically understand the context of the website. In Tab.~\ref{tab:semanticClustersWordsExample}, cluster $4$  contains words related to \textit{e-commerce}, whereas in cluster $5$ words associated with \textit{images} are considered.
\end{itemize}


      

\section{Trials and Experimentation}
The aim of the experiments is to demonstrate that a dirbusting strategy based on semantic clustering enhances the discovery of a website structure, by reducing the number of HTTP requests required to successfully complete the entire process.
To the purpose, we developed a virtualized environment composed of $8$ distinct target websites.
Then, we created a wordlist containing full paths of each website and instrumented a dirbusting tool that can be configured to run either in bruteforce mode or by leveraging the semantic clustering strategy introduced by us. 
In order to create the wordlist, we have started all of the target applications and retrieved full paths by executing OS-level commands.  
In Tab.~\ref{tab:wordCount} the word count for each of the web applications under attack is reported.


    

\begin{table}[H]
    \centering
    \begin{tabular}{ | c | c | c |}
    \hline
    Web Application & Words Count & Total \%\\ \hline\hline
    bodgeit & 40 & 0.47\% \\ \hline
    bricks  & 66 & 0.78\%   \\ \hline
    drupal (CMS) & 1074 & 12.82\%  \\ \hline
    DVWS & 80  & 0.95\% \\ \hline
    Joomla (CMS)  & 4672 & 55.78\% \\ \hline
    Wacko  & 126 & 1.50\%  \\ \hline 
    Wordpress (CMS)  & 1595 & 19.04\% \\ \hline
    XVWA  & 722  & 8.62\% \\ \hline\hline
    Total  & 8375  & 100\% \\ \hline
    \end{tabular}
    \caption{Web applications words count}
    \label{tab:wordCount}
\end{table}  

After obtaining the wordlist, we have applied  semantic clustering in order to group words according to their semantic meaning. The resulting clusters are stored in an ad hoc configuration file that is used by the instrumented dirbusting module when carrying out the clustering-based testing campaigns.

Finally, for each website, we have performed the experiments, by executing the dirbusting tool both in \emph{bruteforce} mode and in \emph{semantic clustering} mode. The results of each experiment have been logged, so as to enable further off-line analysis of the collected data.

The virtualization environment is a useful alternative to a real-world setup for several reasons: 

\begin{itemize} 

\item we do not have to deal with network issues that might affect the environment;
\item we do not create potential Denial Of Service conditions. Indeed, as the dirbusting process sends lots of requests against a web application, if the tested webserver is not designed to support high traffic loads, it might crash;
\item we do not run into legal issues: a bruteforce directory listing might be tagged as a bruteforce attack. Dirbusting is an inner part of Penetration Testing. As such, it should be regulated by contracts. 
\end{itemize}

As described in Section~\ref{section:intro}, we define \textit{``valid request''} an HTTP request that has discovered a new path in the target website. We compare the total number of executed HTTP requests with the number of valid HTTP requests.
To improve the significance of the results, we have repeated each experiment $30$ times for each website.

With the bruteforce approach, the unique wordlist is shuffled and words are used to perform the classic dirbusting procedure. On the other hand, with the clustering approach, we use the algorithm we have described in the previous section to select the words from the wordlist in a non-random way. 

We have compared results graphically, by plotting the relation between the number of total requests sent to the target website and the number of valid requests. 



Our goal is to compare the above mentioned approaches by measuring the overall number of requests sent to the web application in order to thoroughly complete the discovery of a website's structure. The aim of such a comparative evaluation is to show that performance increases when using the semantic clustering approach. 

\subsection{Experiment Information}

The following information is useful to better describe the performed experiments. First, we do not evaluate the time required to complete the task. We rather verify that our solution reduces the number of HTTP requests sent in order to reconstruct the entire structure of the target website. For this reason, we do not 
compare the execution time of the two approaches.

As already anticipated, we have created an isolated environment by using container-based virtualization. No highly-intensive processes have been run during the tests. We continuously monitored CPU, RAM and disk usage during each experiment, so to get sure that none of them went under pressure during any of the trials. We also verified that the tool did not crash during any of the runs.

Network issues might affect the results of the campaign. Even though we do not focus on response times, it is important to have a stable network, as network problems might lead to response timeouts, thus invalidating results. For the above reasons, websites are deployed in an isolated docker network. The environment runs in the context of the dirbusting instrumented tool. We can hence safely assume that there are no network reliability issues during the experiments.

\subsection{Architecture of the benchmarking tool} 

\begin{figure*}[h!]
    \centering
    \includegraphics[width=0.8\linewidth]{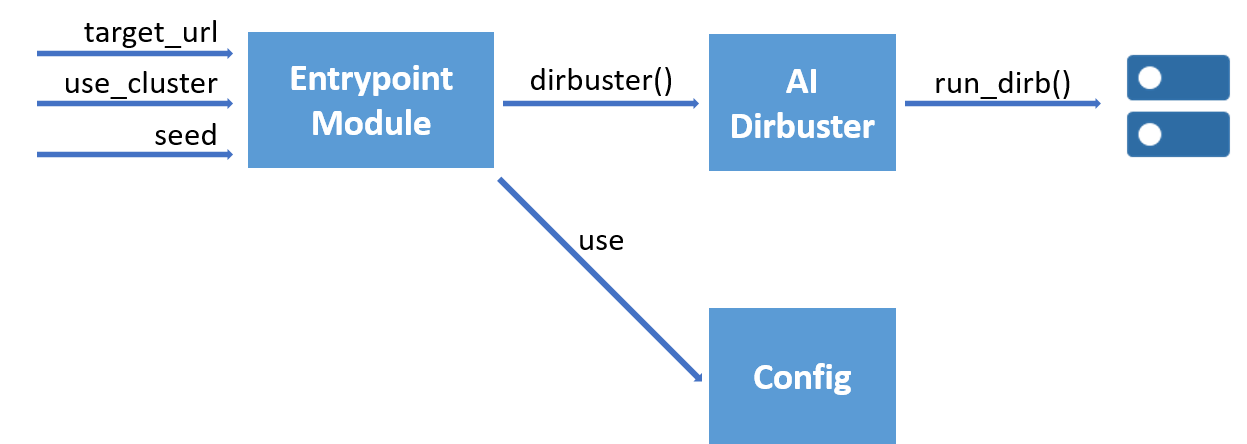}
    \caption{Architecture of the dirbusting Tool}
    \label{fig:dirbustingTool}
\end{figure*}

To perform experiments, we instrumented a dirbusting tool that is illustrated in Figure~\ref{fig:dirbustingTool}

The tool in question was developed in Python 3.6 and is made of three components:

\begin{itemize}
\item \textit{Entrypoint}: accepts input parameters needed to set up a trial associated with a specific target;
\item \textit{AiDirBuster}: dirbusting module that implements the dirbusting process either in bruteforce mode or by leveraging the semantic clustering strategy proposed by us; 
\item \textit{Config}: configuration module containing several configuration parameters, such as groups of words identified through semantic clustering.
\end{itemize}

The tool accepts the following parameters as inputs: 

\begin{itemize}
    \item \textit{use\_clustering}: a boolean value. If true, dirbusting uses the semantic clustering strategy; otherwise, a bruteforce approach is adopted; 
    \item \textit{target\_url}: the target web application used to run the experiment; 
    \item \textit{seed}: a seed used to increase randomness when the wordlist is shuffled during the bruteforce approach. 
\end{itemize}

Semantic clusters are computed offline and subsequently stored in the above mentioned \textit{Config} module. 


 \textit{Universal Sentence Encoder (USE)}~\cite{cer2018universal}, version 4\footnote{\href{https://tfhub.dev/google/universal-sentence-encoder/4}{https://tfhub.dev/google/universal-sentence-encoder/4}}, implemented in TensorFlow 2.2.0~\cite{tensorflow2015-whitepaper}, is used to extract sentence embeddings.

To find and collect clusters, we leveraged the K-means clustering algorithm implementation made available by the \emph{sklearn} python library. The default set of parameter values was used, with the exception of the factor `K', that was properly configured with the elbow method. 

\subsection{Experimental environment setup} 
To simulate the dirbusting process, we have built a docker environment composed of $8$ publicly available web applications, some of which are also typically used for experimenting with vulnerability assessment and penetration testing. 


Table \ref{tab:my_label} shows the characteristics of the web applications in question.
    
\begin{table}[H]
    \centering
    \begin{tabular}{ | c | c | c |}
    \hline
    Web Application & Language Extension & path name convention \\ \hline\hline
    bodgeit & jsp  & under case  \\ \hline
    bricks & php  & camel case   \\ \hline
    drupal (CMS) & php  & snake case   \\ \hline
    DVWS & php   & snake case   \\ \hline
    Joomla (CMS) & php  & upper case \\ \hline
    Wacko & php & snake case   \\ \hline 
    Wordpress (CMS) & php & kebab case  \\ \hline
    XVWA & php & snake case   \\ \hline
    
    \end{tabular}
    \caption{Web applications used for the experiment}
    \label{tab:my_label}
\end{table} 

Among the applications reported in the table, the ones that are usually used to experiment with web application penetration  testing are Bodgeit\footnote{\url{https://github.com/psiinon/bodgeit}}, bricks\footnote{\url{https://sourceforge.net/projects/owaspbricks/}}, DVWS\footnote{\url{https://github.com/snoopysecurity/dvws}} (Damn Vulnerable Web Services), XVWA\footnote{\url{https://github.com/s4n7h0/xvwa}} (Xtreme Vulnerable Web Application) and Wacko\footnote{\url{https://github.com/adamdoupe/WackoPicko}}.

On the other hand, Wordpress\footnote{\url{https://wordpress.com/}}, Drupal\footnote{\url{https://www.drupal.org/}} and Joomla\footnote{\url{https://www.joomla.org/}} are among the most widely spread PHP Content Management Systems used to create web applications. 

\begin{figure}
    \centerline{
    \includegraphics[width=\linewidth]{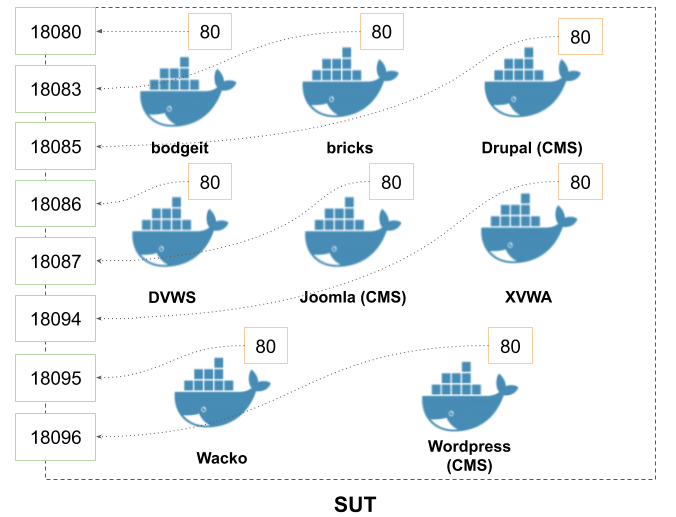}}
    \caption{Container-based testbed architecture}
    \label{fig:experimentArch}
\end{figure}

The environment realized for the experiment is built by using an Infrastructure as Code (IaC) approach. A docker-compose file including eight services describes the architecture of the system under test, as shown in Fig.~\ref{fig:experimentArch}. Each service exposes the standard HTTP service (port $80$) and maps it onto an unassigned TCP port of the hosting machine. The semantic clustering dirbusting tool sends requests to the eight web applications by targeting such exposed TCP ports on the host. With this approach, it is possible to add new web applications in an easy way, as well as to extend the experiment by including new target applications. 

\subsection{Wordlist acquisition}

To create the integrated wordlist, we extracted paths from each webserver and merged them in a single file. Given \textit{n} the number of webservers used for the experiments, the following formula applies: 

\[UniqueWordlist = rand(\sum_{w=1}^{n} absolute\_paths_{w})\]

In a nutshell, \textit{UniqueWordlist} can be obtained as the randomized concatenation of all absolute paths contained in each webserver. 

The path extraction task for a specified webserver can be carried out by executing OS-level commands inside the related Docker service. The following one-line command is a practical example of how the above mentioned task might be completed:

\begin{small}
\begin{verbatim}
docker exec -it <webserver> bash 
cd /var/www 
find .  | sed 's/^\.//g'
\end{verbatim}
\end{small}

The concatenation of all of the collected words creates a unified wordlist. As described in the previous section, the words in the wordlist are basically absolute paths. For our experiments, the final unified wordlist is composed of $8367$ words.

\section{Experimental Results}

The experiments we have conducted allowed us to demonstrate that the enhanced dirbusting strategy we propose actually outperforms the legacy bruteforce approach. Indeed, for each of the eight web servers under test we were able to achieve a performance improvement that is up to $50$\%.

\begin{figure*}%
\centering
\begin{subfigure}{.49\textwidth}
\includegraphics[width=\columnwidth]{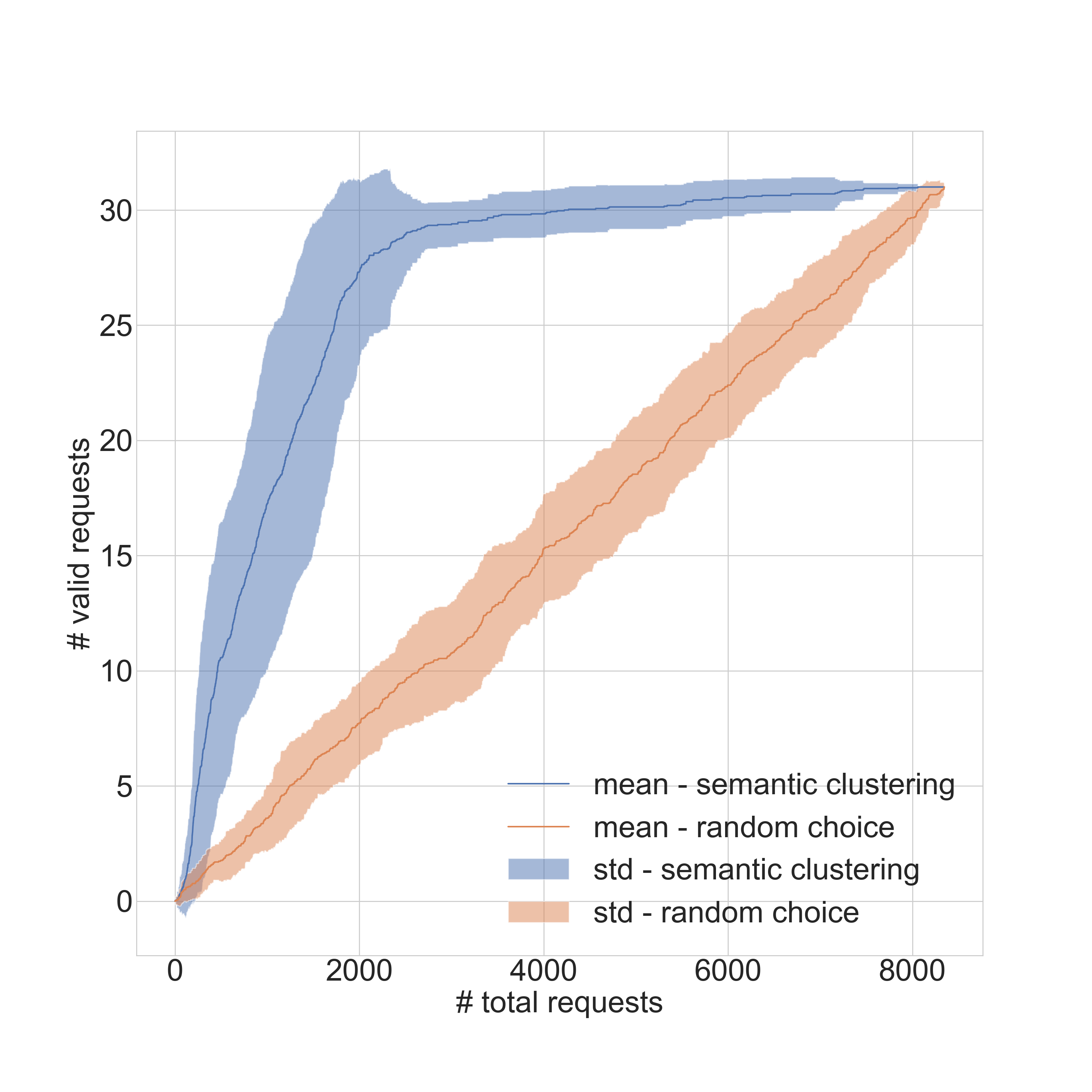}%
\caption{Bodgeit}%
\label{subfiga}%
\end{subfigure}\hfill%
\begin{subfigure}{.49\textwidth}
\includegraphics[width=\columnwidth]{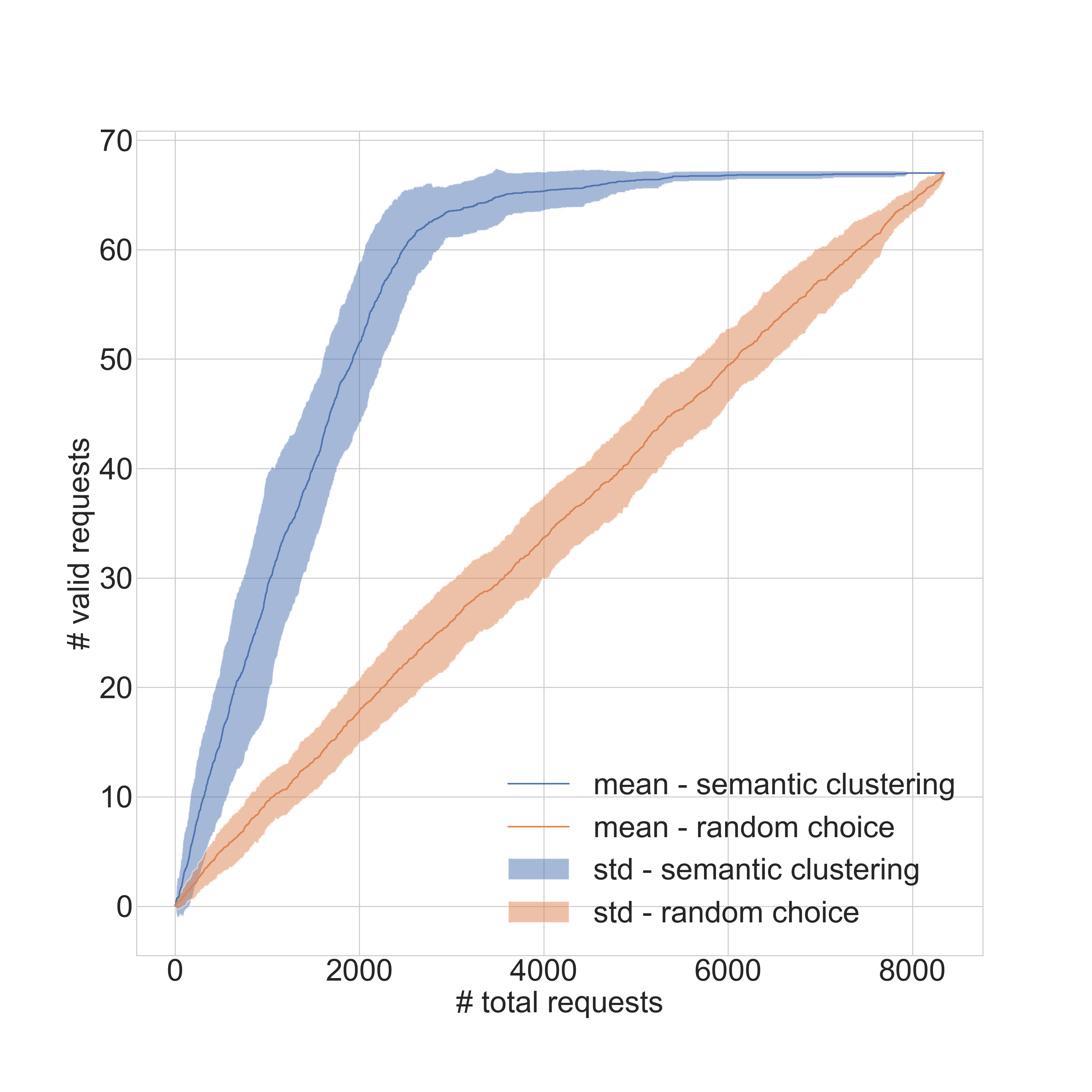}%
\caption{Bricks}%
\label{subfigb}%
\end{subfigure}\hfill%

\begin{subfigure}{.49\textwidth}
\includegraphics[width=\columnwidth]{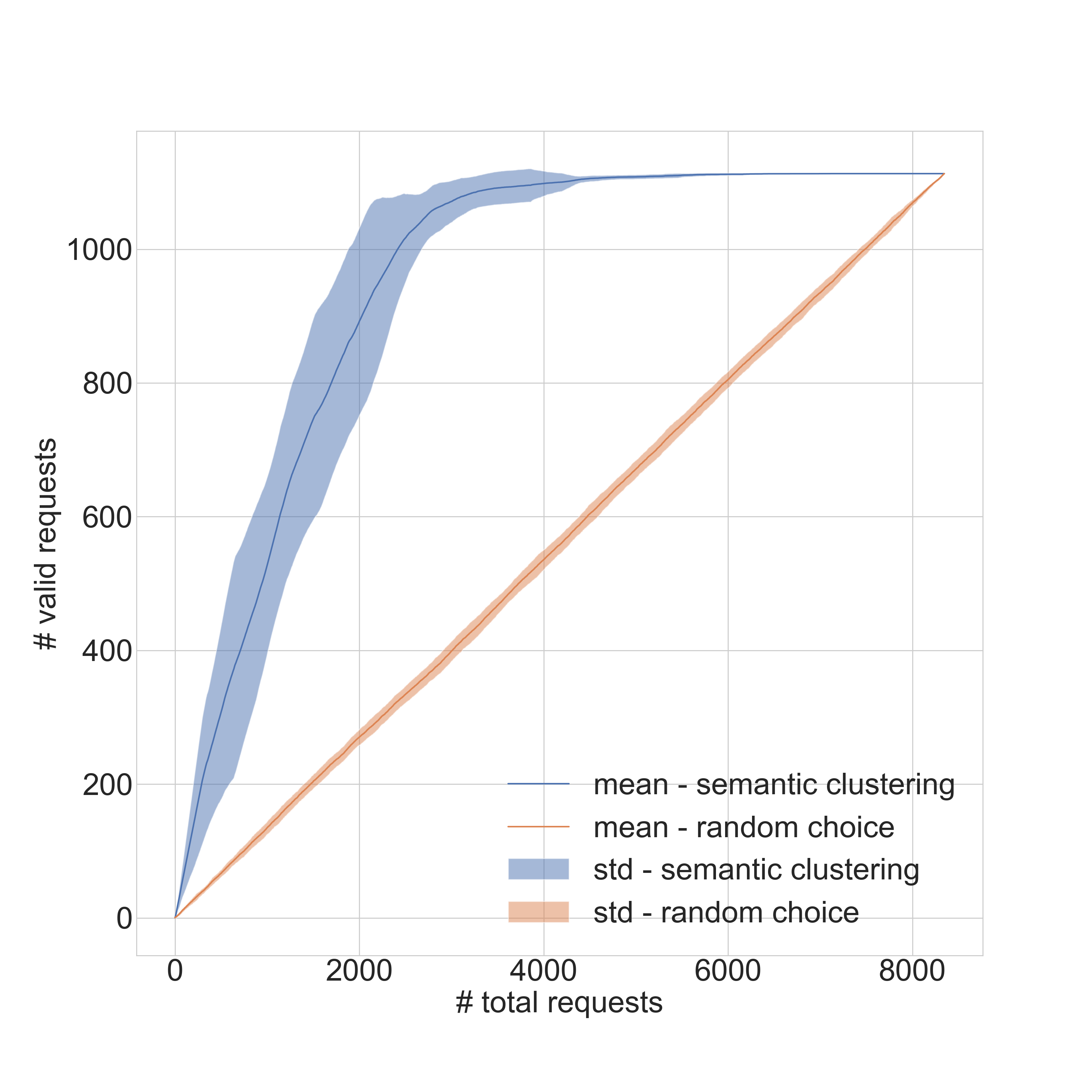}%
\caption{Drupal}%
\label{subfiga}%
\end{subfigure}\hfill%
\begin{subfigure}{.49\textwidth}
\includegraphics[width=\columnwidth]{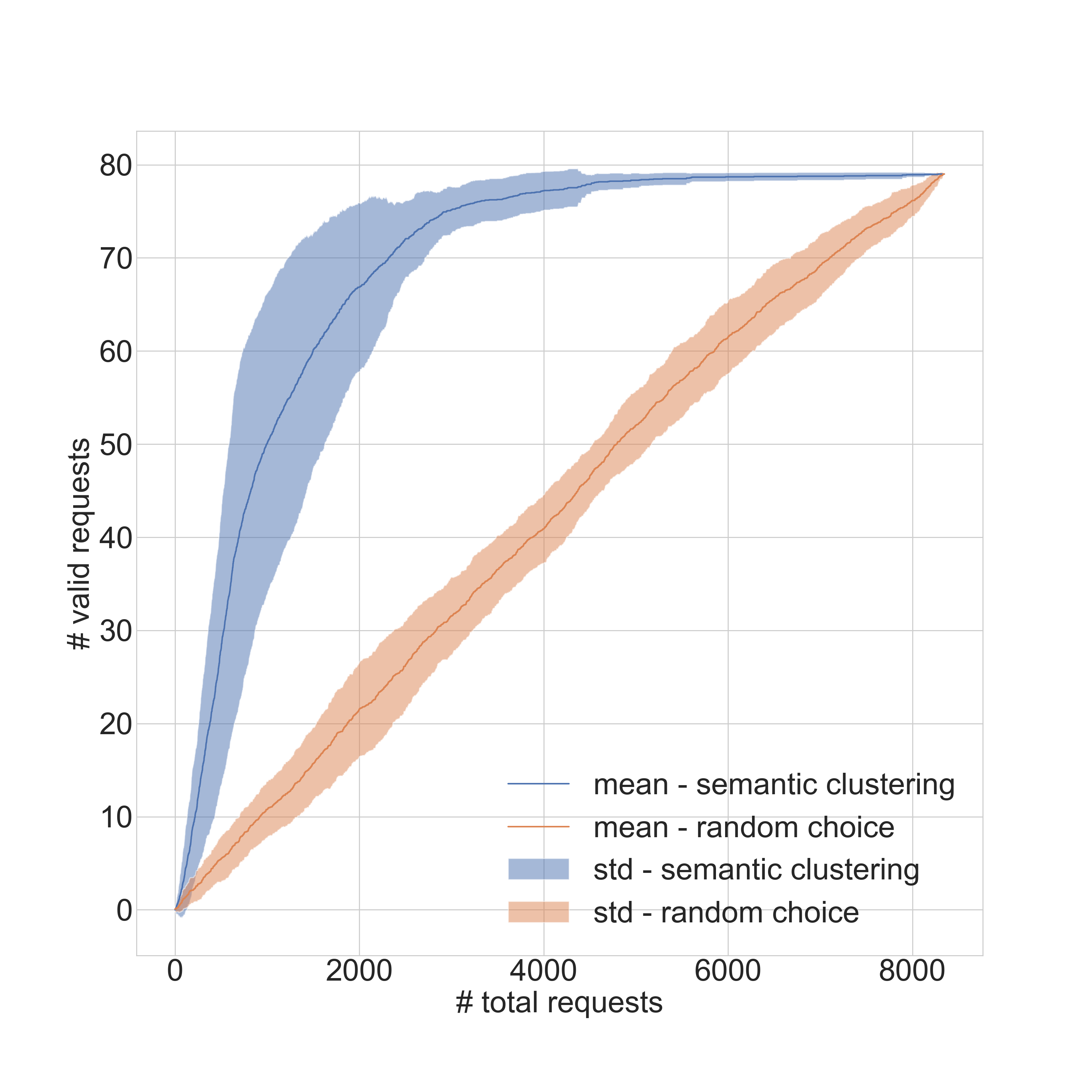}%
\caption{DVWS}%
\label{subfigb}%
\end{subfigure}\hfill%

\caption{Performance plots - 1}
\label{fig:experiments1}
\end{figure*}

\begin{figure*}%
\centering

\begin{subfigure}{.49\textwidth}
\includegraphics[width=\columnwidth]{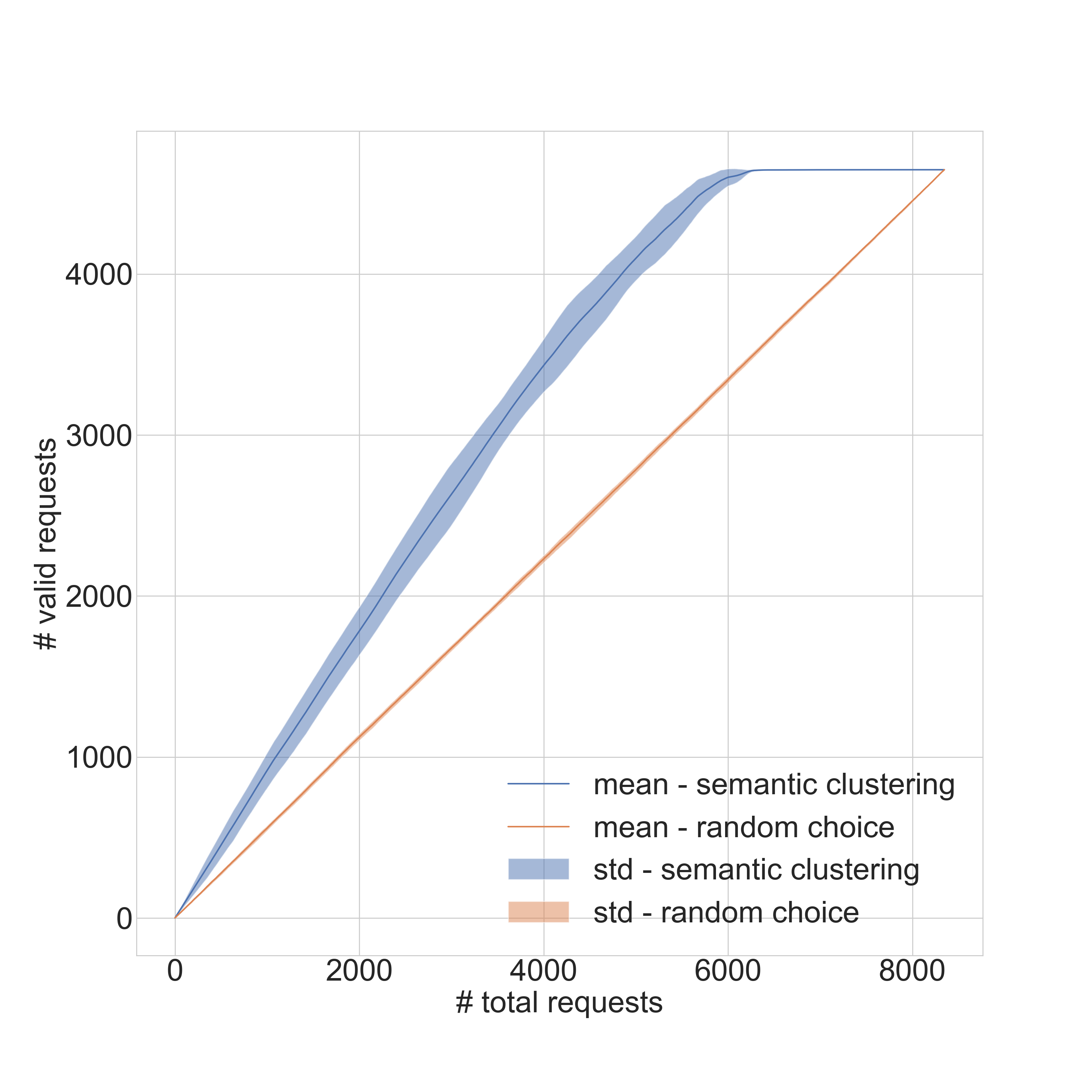}%
\caption{Joomla}%
\label{subfiga}%
\end{subfigure}\hfill%
\begin{subfigure}{.49\textwidth}
\includegraphics[width=\columnwidth]{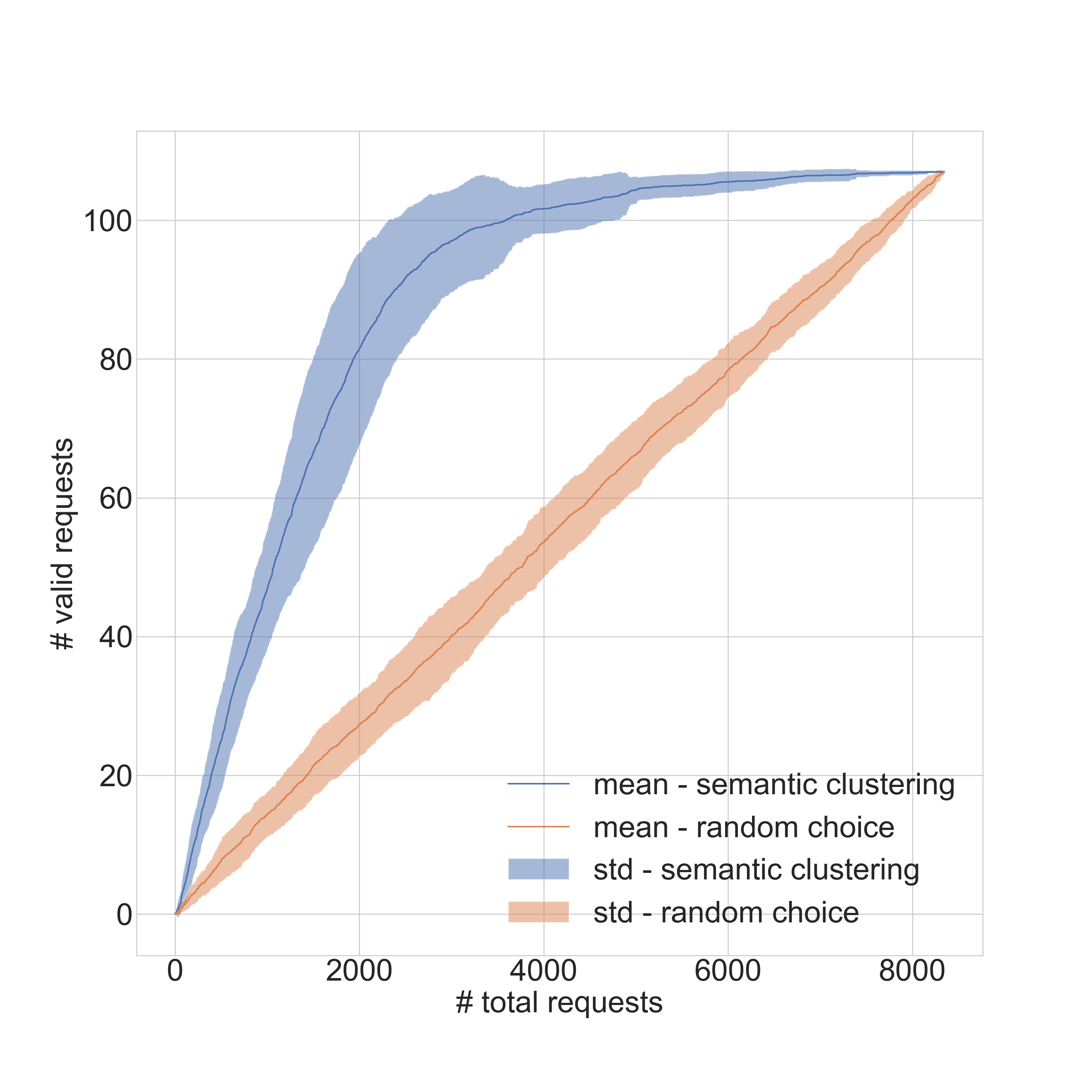}%
\caption{Wacko}%
\label{subfigb}%
\end{subfigure}\hfill%

\begin{subfigure}{.49\textwidth}
\includegraphics[width=\columnwidth]{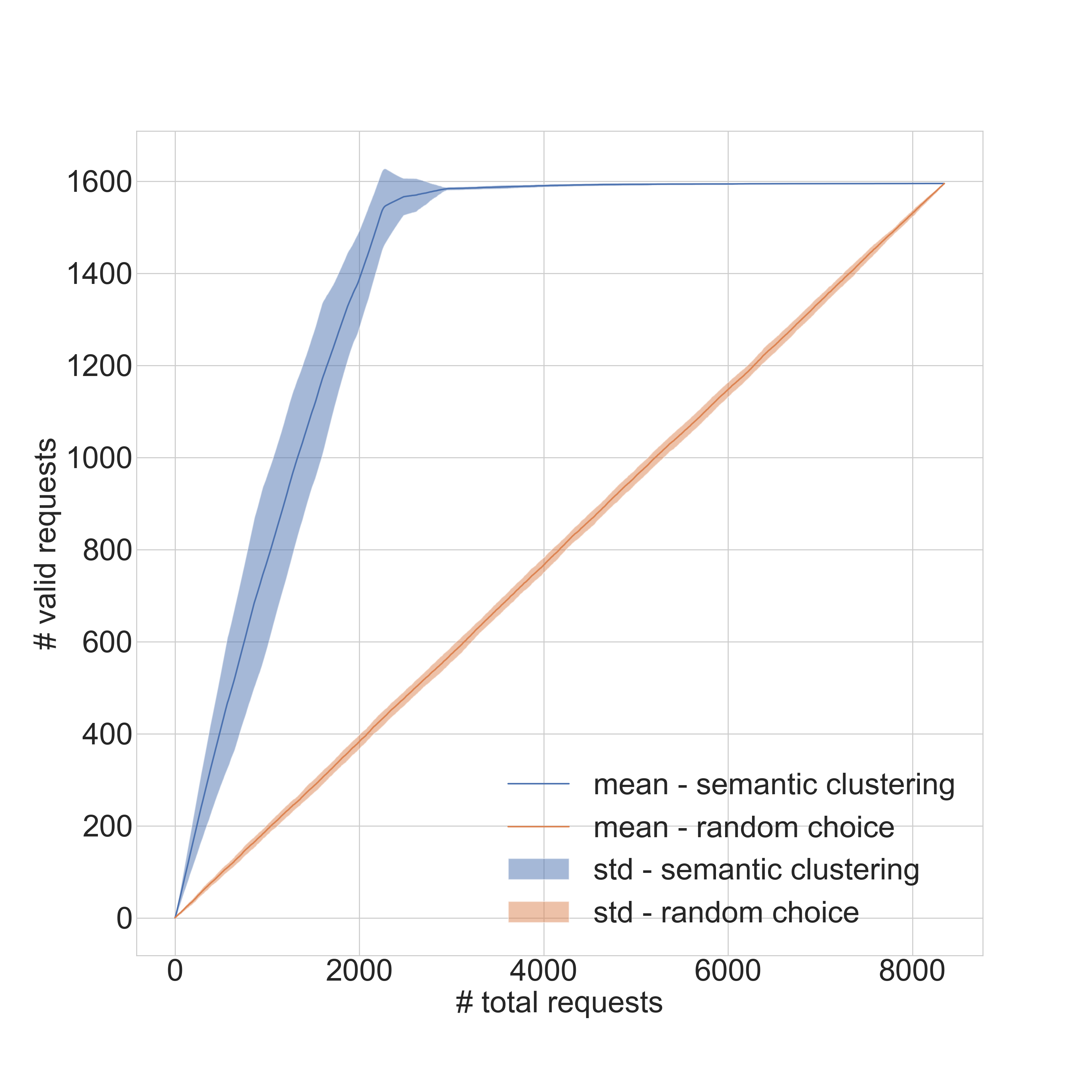}%
\caption{Wordpress}%
\label{subfiga}%
\end{subfigure}\hfill%
\begin{subfigure}{.49\textwidth}
\includegraphics[width=\columnwidth]{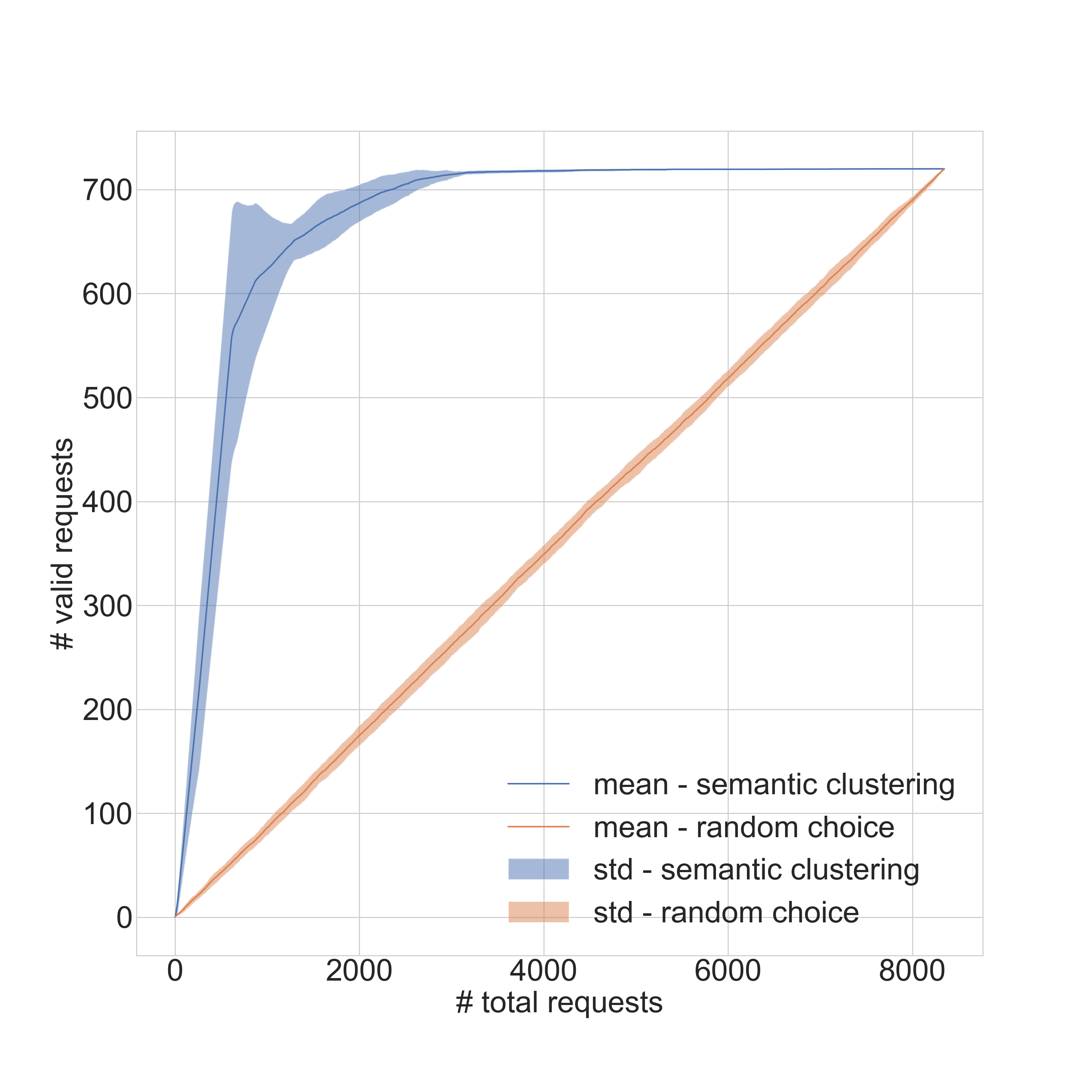}%
\caption{XVWA}%
\label{subfigb}%
\end{subfigure}\hfill%

\caption{Performance plots - 2}
\label{fig:experiments2}
\end{figure*}

In Fig.~\ref{fig:experiments1} and Fig.~\ref{fig:experiments2} we show the results of our campaign. For each web server, we plot both the mean and the standard deviation (\emph{std}) trend of the detected valid requests, over the number of total requests addressed to the target server. Each experiment has been replicated $30$ times in order to improve the significance of the collected results.

As we anticipated above, a ``valid request'' is a request with response code other than 404 (\emph{Not Found} HTTP error message). 

In each of the plots the two approaches are compared. In orange we show the results of the legacy random bruteforce approach, while in blue we report the performance of the proposed semantic clustering strategy.

As it is possible to observe, the random bruteforce strategy shows a linear trend. Indeed, as we apply randomization for each experiment, the number of requests needed to find all the paths is, on average, equal to the number of paths. In this way, the longer the wordlist, the higher will be the number of requests required to find the valid paths.

On the other hand, semantic clustering shows a steeper growth rate. As a matter of fact, with this approach the curve stops increasing much earlier than with the brute force one. 

With all web servers under test, the trend of the two approaches never overlaps. This clearly indicates the gain in performance that can be achieved by leveraging the proposed semantic clustering approach.

Moreover, our approach is able to detect almost all of the available target URLs with half of the requests in relation to the bruteforce approach, hence providing a performance improvement that is close to $50$\%. 

The only exception is Joomla, where the performance improvements are lower compared to the other web applications under test. 
The reason behind such a finding is that the number of words collected for Joomla is $4672$, that is more than half the total number of the words included in our wordlist. This is clear by looking at the words count in Table~\ref{tab:wordCount}. As it is possible to observe, Joomla covers about half of the integrated wordlist.
This entails that with a random approach we are able to find valid URLs, in Joomla, with a probability of about $50$\%, hence reducing the gain that can be achieved by leveraging the alternative approach proposed by us. Nevertheless, as it is possible to observe, there is still a clear improvement for this web application as well, with around $2000$ requests less than those needed to find all of the available paths with the bruteforce approach.

\section{Conclusions}
In this work we illustrated how it is possible to improve dirbusting by leveraging Semantic Clustering.
Our approach has been proven to be effective on $8$ different web applications, where the clustering approach significantly improved performance in each of the conducted experiments. 

In future works we are going to further improve the effectiveness of our approach by combining it with other types of techniques. As an example, we will properly combine standard web spidering with dirbusting. We will first leverage spidering to detect the overall structure of the target web application. Then, we will trigger dirbusting in order to dig deeper and discover hidden or private pages.



We also remark that our work is focused on demonstrating that a semantic clustering approach performs better than a bruteforce one to discover the web application structure.  For this reason, we do not consider strategies to explore subpaths. Wordlists used during the experimentation are therefore composed by full paths (e.g., \textit{/users/mooney/ir-course/}). Though, dirbusting techniques can be recursive. Namely, whenever a new path is found, dirbusting can be recursively applied to it in order to discover new sub-paths. 
In view of the above considerations, it would be interesting to investigate  the use of our semantic clustering approach in a recursive way, while also evaluating the possible alternative strategies for applying recursion while navigating through the dynamically identified sub-paths (e.g., breadth-first, depth-first, etc.).

\newpage

\bibliographystyle{unsrt} 
\end{document}